\pdfoutput=1
\documentclass[11pt,twoside,a4paper,cmspaper,final,collab]{cms-tdr}
\def\svnVersion{b3650be-D}\def\svnDate{2021/05/28}\def\cmsCernNoTag{CERN-EP-2020-226}\def\cmsCernDate{\today}\def\cmsMessage{"Published in the Journal of High Energy Physics as \href{http://dx.doi.org/10.1007/JHEP05(2021)284}{\doi{10.1007/JHEP05(2021)284}.}"}
\begin{document}\cmsNoteHeader{HIN-18-014}

\newcommand {\ncoll}    {\ensuremath{N_{\text{coll}}}}
\newcommand {\TAA}      {\ensuremath{\langle T_{\mathrm{AA}}\rangle}\xspace}
\newcommand {\ak}       {anti-\kt\xspace}
\newcommand {\RAA}      {\ensuremath{R_{\mathrm{AA}}}\xspace}
\newcommand {\RAARatio} {\ensuremath{R_{\mathrm{AA}}^R/R_\mathrm{AA}^{R=0.2}}\xspace}

\newcommand {\pp}       {\ensuremath{\Pp\Pp}\xspace}
\newcommand {\PbPb}     {\ensuremath{\text{PbPb}}\xspace}

\newcommand {\ptj}      {\ensuremath{\pt^\text{jet}}\xspace}
\newcommand {\etaj}     {\ensuremath{\eta^\text{jet}}\xspace}

\newcommand{\PYTHIAHYDJET} {\PYTHIA{+}\HYDJET\xspace}
\newcommand{\BDMPS} {\textsc{bdmps}\xspace}
\newcommand{\Martini} {\textsc{Martini}\xspace}
\newcommand{\SCET} {\textsc{Scet}\xspace}
\newcommand{\LBT} {\textsc{lbt}\xspace}
\newcommand{\CCNU} {\textsc{Ccnu}\xspace}
\newcommand{\Hybrid} {\textsc{Hybrid}\xspace}
\newcommand{\pyquen} {{\textsc{pyquen}}\xspace}
\newcommand{\jewel} {{\textsc{Jewel}}\xspace}

\cmsNoteHeader{HIN-18-014} 

\title{First measurement of large area jet transverse momentum spectra in heavy-ion collisions}

\date{\today}

\abstract{
Jet production in lead-lead (\PbPb) and proton-proton (\pp) collisions at a nucleon-nucleon center-of-mass energy of 5.02\TeV is studied with the CMS detector at the LHC, using \PbPb and \pp data samples corresponding to integrated luminosities of 404\mubinv and 27.4\pbinv, respectively. Jets with different areas are reconstructed using the \ak algorithm by varying the distance parameter $R$. The measurements are performed using jets with transverse momenta (\pt) greater than 200\GeV and in a pseudorapidity range of $\abs{\eta}<2$. To reveal the medium modification of the jet spectra in \PbPb collisions, the properly normalized ratio of spectra from \PbPb and \pp data is used to extract jet nuclear modification factors as functions of the \PbPb collision centrality, \pt and, for the first time, as a function of $R$ up to 1.0. For the most central collisions, a strong suppression is observed for high-\pt jets reconstructed with all distance parameters, implying that a significant amount of jet energy is scattered to large angles. The dependence of jet suppression on $R$ is expected to be sensitive to both the jet energy loss mechanism and the medium response, and so the data are compared to several modern event generators and analytic calculations. The models considered do not fully reproduce the data.
}

\hypersetup{
pdfauthor={CMS Collaboration},
pdftitle={First measurement of large area jet transverse momentum spectra in heavy-ion collisions},
pdfsubject={CMS},
pdfkeywords={CMS, heavy-ion collision, jets}}

\maketitle
\section{Introduction}
Quantum Chromodynamics (QCD), the theory of the strong nuclear force, predicts that a deconfined state of quarks and gluons, called the quark-gluon plasma (QGP), should be produced at sufficiently high temperatures and densities~\cite{Karsch:1995sy}.
In relativistic heavy ion collisions, the QGP is produced on an extremely short time scale~\cite{PhysRevD.27.140,Bernhard:2016tnd}.
A pair of partons (quarks or gluons) in the colliding nuclei can undergo a high transverse momentum (\pt) scattering, a process that occurs prior to the formation of the QGP.
As the scattered partons pass through and interact with the QGP, they lose some of their energy, thereby acting as probes of the short-distance structure of the medium~\cite{Bjorken:1982tu,Gyulassy:1990ye,Wang:1991xy,Baier:1996sk,Zakharov:1997uu}. This parton energy loss, often referred to as ``jet quenching'', is related to the transport and thermodynamical properties of the QGP~\cite{Appel:1985dq,Blaizot:1986ma,Burke:2013yra,Liu:2006ug}.
However, the details of the parton's interactions with the medium, as well as the relative importance of each interaction mechanism, are not yet fully understood~\cite{CasalderreySolana:2007pr, d'Enterria:2009am, Wiedemann:2009sh, Majumder:2010qh,Qin_2015,cao2020jet}.

A hard-scattered parton fragments and hadronizes into a collimated spray of particles.  The fragmentation process coevolves with the QGP.
The suppression of inclusive high-\pt hadrons in nucleus-nucleus collisions~\cite{Adcox:2004mh,Adams:2005dq,Back:2004je,Arsene:2004fa,Adare:2007vu,CMS:2012aa,Aad:2015wga,Aamodt:2010jd} provides evidence for jet quenching.
Experimentally, final-state particles can be clustered into jets through the use of well-defined algorithms such as \ak~\cite{Cacciari:2008gp}.
Various studies of jets and jet pairs, such as dijet \pt imbalance~\cite{Aad:2010bu,Chatrchyan:2011sx,Adamczyk_2017}, modifications of the jet yield in the medium~\cite{Aad:2014bxa, Adam:2015ewa, Khachatryan:2016jfl,Adam_2020,ALICERAA2020,ATLASRAA2019,ALICERAA2015}, electroweak boson-jet \pt imbalance~\cite{Chatrchyan:2012gt, Sirunyan:2017qhf}, jet fragmentation functions~\cite{Chatrchyan:2014ava, Aaboud:2017bzv,Sirunyan:2018qec}, missing \pt in dijet systems~\cite{Aad:2010bu, Chatrchyan:2011sx,
Khachatryan:2015lha}, jet-track correlations~\cite{Khachatryan:2016tfj}, and the radial \pt profile of tracks within jets~\cite{Chatrchyan:2013kwa,CMS-HIN-16-020,CMS-HIN-18-006,ALICEJetRadialProfile,StarHJet} have been studied.
Complementary to these measurements, inclusive jet spectra reconstructed using different distance parameters $R$ in the \ak algorithm are of great interest because they are less sensitive to hadronization effects than observables involving individual final-state hadrons.
The value of $R$ defines the area of the reconstructed jet.
By varying $R$, different fractions of energy from the quenched jet and the medium response will be included in the reconstructed jet.
A differential study of the suppression versus $R$ provides new sensitivity to the QGP properties~\cite{Chien:2015hda} and to the underlying jet quenching mechanism. In particular, theoretical models and generators based on perturbative QCD~\cite{Armesto:2009fj,Chien:2015hda,Tachibana:2017syd} and anti-de Sitter/conformal field theory correspondence~\cite{Hulcher_2018} predict different dependences of the jet suppression on $R$.

Modifications to jet production can be quantified by the ratio of the inclusive jet yields per event
in nucleus-nucleus (AA) collisions ($N^{\mathrm{AA}}$) to those in proton-proton (\pp) collisions ($N^{\pp}$), scaled by the mean number of binary nucleon-nucleon (NN) collisions ($\langle \ncoll \rangle$)~\cite{Loizides:2017ack}. This ratio is called the nuclear modification factor \RAA and is defined as 
\begin{linenomath}
\begin{equation}
\RAA(\ptj) = \frac{\rd N^\mathrm{AA}/\rd\ptj}{\langle N_\text{coll}\rangle \rd N^{\pp}/\rd\ptj}=
\frac{\rd N^\mathrm{AA}/\rd\ptj}{\TAA\,\rd\sigma^{\pp}_\text{inel}/\rd\ptj}\;,
\label{eq:raa}
\end{equation}
\end{linenomath}
where \ptj is the transverse momentum of the jet.
The \RAA is typically measured in bins of centrality, which characterizes the degree of overlap of the two colliding lead nuclei~\cite{Chatrchyan:2011sx,Miller:2007ri}.
The nuclear overlap function \TAA is defined as the ratio of $\langle \ncoll \rangle$ to the total inelastic \pp cross section, $\TAA = \langle \ncoll \rangle /\sigma^{\pp}_\text{inel}$, and can be calculated from a Glauber model of
the nuclear collision geometry~\cite{Miller:2007ri}.
If the ratio is less than one, it indicates a transfer of energy out of the jet cone.
Measurements of the dependence of jet spectra and nuclear modification factors on the jet distance parameter $R$ can help differentiate between competing models of parton energy loss mechanisms~\cite{pablos2019jet}.

In studies of jet suppression from LHC Run 1 with lead-lead (\PbPb) collisions at a nucleon-nucleon center-of-mass energy of $\sqrtsNN=2.76\TeV$, it was shown that the level of suppression is roughly independent of \ptj in the range \ptj = 200--400\GeV~\cite{Khachatryan:2016jfl}.
This suggests that the shape of the spectra is not significantly modified, and the modifications are predominantly through the overall number of jets.
However, these initial measurements were statistically limited. At $\sqrtsNN=5.02\TeV$, this measurement can be extended to higher \pt. Furthermore, at this higher center-of-mass energy, partons traverse a medium of higher density and temperature. 

In this paper,  measurements of jet \RAA at $\ptj>200\GeV$ using \PbPb collisions at $\sqrtsNN = 5.02\TeV$ are reported. The jets are reconstructed using the \ak algorithm~\cite{Cacciari:2008gp} with $R$ varying between 0.2 and 1.0.
The results are presented as a function of \ptj in bins of \PbPb event centrality.

\section{The CMS apparatus}
The central feature of the CMS detector is a superconducting solenoid of 6\unit{m} internal diameter, providing a magnetic field of 3.8\unit{T}. Within the solenoid volume are a silicon pixel and strip tracker, a lead tungstate crystal electromagnetic calorimeter (ECAL), and a brass and scintillator hadron calorimeter (HCAL), each composed of a barrel and two endcap sections. Hadron forward (HF) calorimeters extend the pseudorapidity coverage up to $\abs{\eta}=5.2$ and are used for event selection. In the case of \PbPb events, the HF signals are also used to determine the centrality class of the event.
In the barrel section of the ECAL, an energy resolution of about 1\% is achieved for unconverted or late-converting photons that have energies in the range of tens of GeV. The remaining barrel photons have a resolution of about 1.3\% up to $\abs{\eta} = 1$, rising to about 2.5\% at $\abs{\eta} = 1.4$. In the endcaps, the resolution of unconverted or late-converting photons is about 2.5\%, while the remaining endcap photons have a resolution between 3 and 4\%~\cite{CMS:EGM-14-001}. When combining information from the entire detector, the jet energy resolution amounts typically to 15\% at 10\GeV, 8\% at 100\GeV, and 4\% at 1\TeV, to be compared to about 40, 12, and 5\% obtained when the ECAL and HCAL calorimeters alone are used \cite{Sirunyan:2017ulk}.
Muons are detected in gas-ionization chambers embedded in the steel flux-return yoke outside the solenoid.
The silicon tracker measures charged particles within $\abs{\eta} < 2.5$. It consists of 1440 silicon pixel and 15\,148 silicon strip detector modules. For nonisolated particles of $1 < \pt < 10\GeV$ and $\abs{\eta} < 1.4$, the track resolutions are typically 1.5\% in \pt and 25--90 (45--150)\mum in the transverse (longitudinal) impact parameter \cite{TRK-11-001}.
Events of interest are selected using a two-tiered trigger system~\cite{Khachatryan:2016bia}.
The first level, composed of custom hardware processors, uses information from the calorimeters and muon detectors to select events at a rate of around 100\unit{kHz} within a time interval of less than 4\mus. The second level, known as the high-level trigger, consists of a farm of processors running a version of the full event reconstruction software optimized for fast processing, and reduces the event rate to around 1\unit{kHz} before data storage.
A more detailed description of the CMS detector, together with a definition of the coordinate system used and the relevant kinematic variables, can be found in Ref.~\cite{Chatrchyan:2008zzk}.

\section{Event selection}

The event samples are recorded with dedicated jet triggers with different \ptj thresholds, the smallest of which is $\ptj>80\GeV$~\cite{Sirunyan:2017qhf}.
The efficiencies of the triggering algorithms are evaluated in data and are found to reach unity in both \pp and \PbPb collisions for jets considered in this paper ($\ptj > 200\GeV$).
A number of requirements are made to the events to remove non-collision events (\eg, beam-gas interactions) and to select only inelastic hadronic collisions~\cite{Khachatryan:2016odn,Sirunyan:2017qhf}.
Both \pp and \PbPb events are required to have at least one reconstructed primary interaction vertex with a distance from the center of the nominal interaction region of less than 15\cm along the beam direction. In addition, in \PbPb collisions the shapes of the clusters in the pixel detector have to be compatible with those produced by a genuine collision~\cite{Khachatryan:2010xs}. 
The \PbPb collision events are also required to have at least three towers in each of the HF detectors with energy deposits of more than 3\GeV per tower. These criteria select $99$\% of inelastic hadronic \PbPb collisions~\cite{Chatrchyan:2011sx}.

The collision centrality for \PbPb events is determined using the total sum of transverse energy from the calorimeter towers in the HF region. The transverse energy distribution is used to divide the event sample into bins of percentage of the total hadronic interaction cross section~\cite{Chatrchyan:2011sx}.
The results in this paper are presented in four centrality intervals, where 0\% corresponds to a full overlap of the two nuclei: 0--10, 10--30, 30--50, and 50--90\%.  The corresponding \TAA and {$\langle\ncoll\rangle$} values used in this paper for the centrality intervals are listed in Table~\ref{tab:TAA}.

\begin{table}[tbh]
\centering
\topcaption{The values of {$\langle\ncoll\rangle$} and \TAA, and their
uncertainties in $\sqrtsNN=5.02\TeV$ \PbPb collisions for the
   centrality ranges used in this analysis~\cite{Loizides:2017ack}.}
\newcolumntype{x}{D{,}{\text{--}}{2.3}}
\newcolumntype{z}{D{,}{}{5.5}}
\setlength\extrarowheight{1.5 pt}
\begin{tabular}{xzz}
\hline
 \multicolumn{1}{c}{Centrality} &  \multicolumn{1}{c}{$\langle \ncoll\rangle$} &  \multicolumn{1}{c}{$\TAA$ [mb$^{-1}$]} \\
\hline
0,10\% & 1630,{^{+120}_{-120}} & 23.2,{^{+0.4}_{-0.7}} \\[1.5pt]
10,30\% & 805,{^{+55}_{-58}} & 11.5,{^{+0.3}_{-0.4}} \\[1.5pt]
30,50\% & 267,{^{+20}_{-20}} & 3.82,{^{+0.21}_{-0.21}} \\[1.5pt]
50,90\% & 30.8,{^{+3.5}_{-2.4}} & 0.440,{^{+0.049}_{-0.032}} \\[1.5pt]
\hline
\end{tabular}
\label{tab:TAA}
\end{table}

\section{Monte Carlo simulations}
Several Monte Carlo (MC) simulated jet event samples are used to evaluate background components, efficiencies, misreconstructed jet rates (arising from upward fluctuations of the underlying event (UE) without a corresponding hard parton), jet energy corrections and jet energy resolutions (JER).
Proton-proton collisions are generated with \PYTHIA 8.212~\cite{Sjostrand:2014zea}, with the UE tune CUETP8M1~\cite{Khachatryan:2015pea}, as well as with \PYTHIA{6}~\cite{Sjostrand:2006za}, with the UE tune Z2~\cite{Chatrchyan_2013} with PDF set CTEQ6L1~\cite{Pumplin_2002}.
For the \PbPb MC samples, each \PYTHIA signal event is embedded into a \PbPb collision event generated with \HYDJET~v1.8~\cite{Lokhtin:2005px}, which is tuned to reproduce global event properties such as the charged-hadron \pt spectrum and particle multiplicity.
The detailed simulation of the CMS detector response is performed using the \GEANTfour package~\cite{geant4}.

\section{Analysis method}

\subsection{Jet reconstruction and underlying event subtraction}

Particle candidates are reconstructed with the particle-flow (PF) algorithm~\cite{Sirunyan:2017ulk}, where information from different parts of the detector are combined to form an optimized description of the event.  Jets are clustered from the PF candidates using the \ak algorithm with distance parameters of $R = 0.2, 0.3, 0.4, 0.6, 0.8$, and 1.0, as implemented in the \FASTJET framework~\cite{Cacciari:2008gp,Cacciari:2011ma}.

One of the main challenges to reconstructing jets in heavy-ion collisions is the additional soft UE coming from the QGP.  In order to subtract the soft UE in \PbPb collisions on an event-by-event basis, an iterative algorithm~\cite{Kodolova:2007hd} is employed. The mean value, $\langle E_{\mathrm{PF}}\rangle$, and dispersion, $\sigma(E_{\mathrm{PF}})$, of the transverse energies from the PF candidates are calculated in a number of $\eta$ bins ~\cite{Chatrchyan:2011sx,Chatrchyan:2012nia,Chatrchyan:2012gt} for each event.
Then, a two-step procedure is employed to account for the azimuthal modulation of background activity arising from the bulk properties of the QGP.
In the first step, the so-called event plane angles ($\Phi_{\mathrm{EP},2}, \Phi_{\mathrm{EP},3}$) for the second- and third-order harmonics of the azimuthal distribution are derived from the HF calorimeters ($3 < \abs{\eta} < 5$)~\cite{Chatrchyan:2012ta}.
This method of estimating the UE gives underlying energy estimations that are consistent with a previous analysis of photon- and {\PZ}-tagged jets in which event plane mixing was used~\cite{CMS-PAPERS-HIN-15-013}.
The event plane angles are not corrected for detector effects since the only goal of this procedure is to obtain a better description of the modulation of the background level.
For the second step, a fit to the azimuthal angle ($\phi$, in radians) distribution of charged-hadron PF candidates with $0.3 < \pt < 3.0\GeV$ and $\abs{\eta} < 1$ is performed.
No explicit exclusion of regions close to the jet is performed, since their effect on the extracted parameters is negligible.  The functional form of the fit is as follows:
\begin{linenomath}
\begin{equation}
  \label{eqn:flowFitForm}
  N(\phi) = N_{0} (1 + 2v_{2}\cos(2[\phi - \Phi_{\mathrm{EP},2}]) + 2v_{3}\cos(3[\phi - \Phi_{\mathrm{EP},3}])),
\end{equation}
\end{linenomath}
where $N_0$ is the magnitude of average UE activity.  The parameters $v_2$ and $v_3$ quantify the strengths of the collective behaviors of the UE known as ``elliptic'' and ``triangular'' flow, respectively.  The event plane angles $\Phi_{\mathrm{EP},2}$ and $\Phi_{\mathrm{EP},3}$ are fixed to the result from the first step.
A fit is performed per event to extract the parameters $N_{0}$, $v_{2}$, and $v_{3}$.
The fit is discarded if the minimum required number of candidates (at least 10 entries in each bin) are not met, or if the reduced $\chi^{2}$ of the fit is greater than 2.
In this case, the background energy density is estimated as a flat distribution in $\phi$, without flow modulations.

\begin{figure}[h!t]
\centering
        \includegraphics[width=0.8\textwidth]{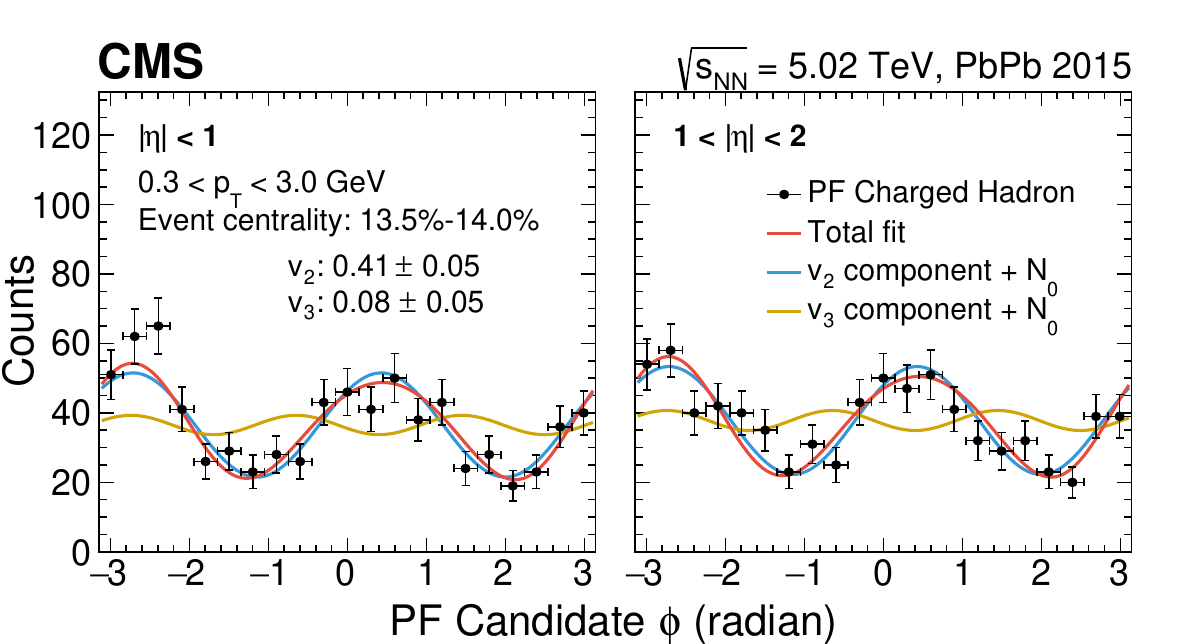}
       \caption{(Color online) Azimuthal angle distributions for a single \PbPb event: $\phi$ modulations at mid-rapidity $\abs{\eta}<1$ (left) and forward rapidity $1<\abs{\eta}<2$ (right) of charged-hadron PF candidates. The $v_2$ (blue curve) and $v_3$ (yellow curve) of the flow components are shown, together with the total modulation used in the analysis to account for the background (red curve).  The flow coefficients are extracted from the left plot and overlaid in the right plot.}
       \label{fig:rhoFlowFitData}
\end{figure}

An example of this procedure is shown for data in Fig.~\ref{fig:rhoFlowFitData}. The left plot shows the fit in the extraction region, along with a breakdown of the components of the fit. The right plot takes parameters extracted from mid-rapidity ($\abs{\eta} < 1$) and renormalizes the function to data at forward-rapidity ($1 < \abs{\eta} < 2$).  A good agreement of the shape for background modulations in the two $\eta$ ranges is observed.

Finally, the UE subtraction in \PbPb collisions is performed using a constituent subtraction method~\cite{Berta:2014eza}. This is a  particle-by-particle approach
that corrects the energy of each jet constituent based on the local average UE density $\rho(\eta, \phi)$. This density is assumed to factorize in $\eta$ and $\phi$ according to the form 
\begin{linenomath}
\begin{equation}
    \rho(\eta, \phi) = \rho(\eta) (1 + 2v_{2}\cos(2[\phi - \Phi_{\mathrm{EP},2}]) + 2v_{3}\cos(3[\phi - \Phi_{\mathrm{EP},3}])). 
    \label{eqn:PhiModulation}
\end{equation}
\end{linenomath}
Here $\rho(\eta)$ encodes the variation of the UE density as a function of $\eta$, and the flow parameters are determined in the previous fit.  The average UE density $\rho(\eta)$ is calculated as the average energy in given $\eta$ bins, excluding regions overlapping with jets.
In \pp collisions, where the UE level is negligible, jets are reconstructed without UE subtraction.

\subsection{Jet energy scale and resolution}

Jet energy corrections are derived from simulation separately for \pp and \PbPb data following methods outlined in Ref.~\cite{Khachatryan:2016kdb}.
The energy scales are verified with an energy balance method applied to dijet and photon+jet events in \pp data. For this study, jets with $\abs{\etaj} < 2$ and (corrected) $\ptj > 160\GeV$ are selected.

\begin{figure}[h!t]
\centering
\includegraphics[width=0.8\textwidth]{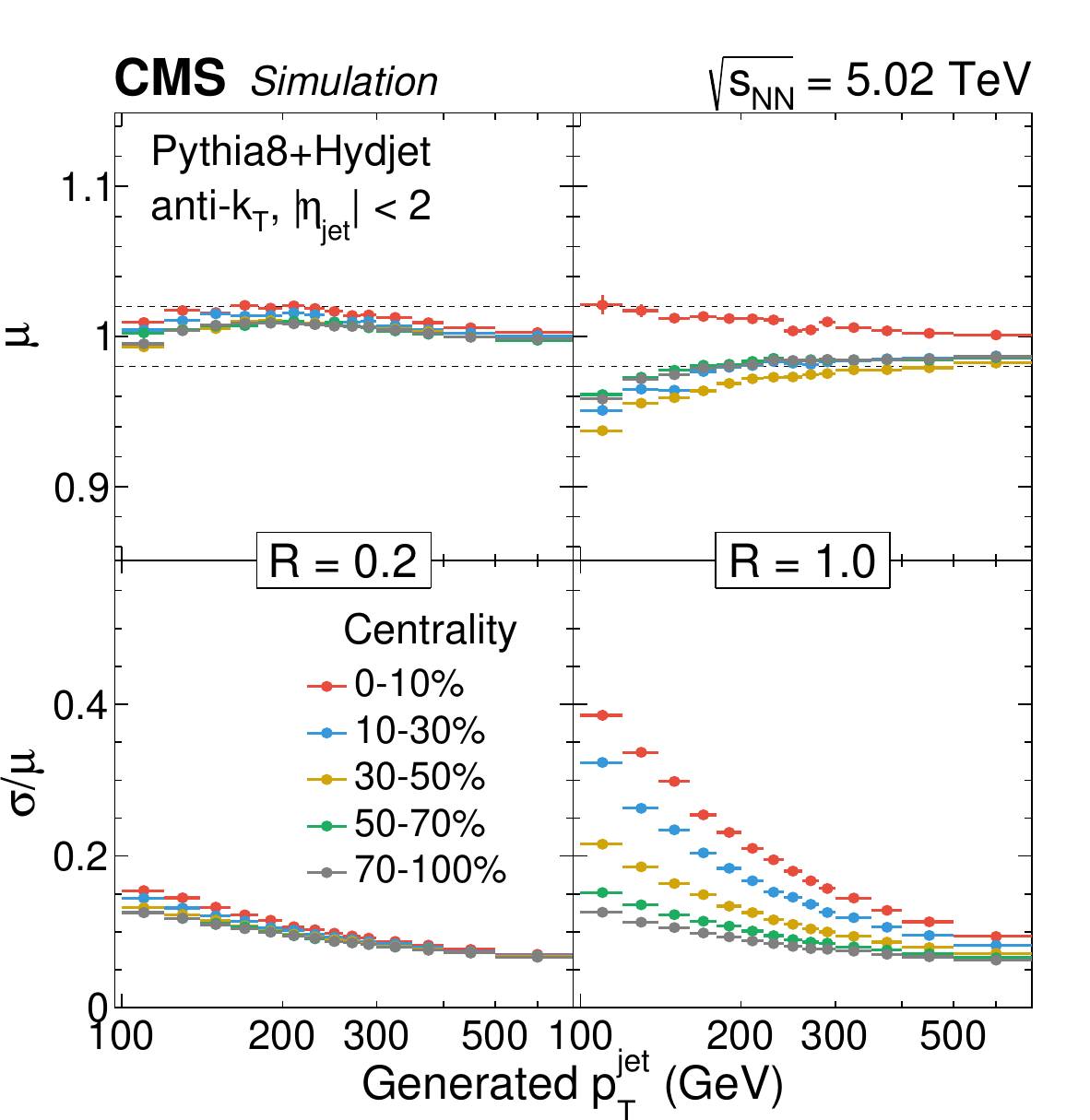}
\caption{The energy scale $\mu$ (upper) and resolution $\sigma/\mu$ (lower) for \PbPb \ak jets with $\abs{\etaj}<2$,  
as functions of 
generated  \ptj. The left (right) column shows jets with  $R=0.2$ (1.0). Several different centrality classes are shown. }
\label{fig:jesJer}
\end{figure}

 The \ptj binning of the analysis is chosen based 
 on the jet energy resolution (JER) for each cone size and centrality. 
For \pp events, the JER varies by less than $10\%$ for different values of $R$. 
These variations reflect how the  probability for energy to move into or out of the jet cone changes with cone sizes.
Figure~\ref{fig:jesJer} shows the \PbPb jet energy scale (JES, upper), defined as the reconstructed \ptj divided by the generated \ptj, and JER (lower), for $R=0.2$ (left) and $R=1.0$ (right) as functions of the generated  \ptj.  The JES is rather flat vs. \ptj while JER decreases with \ptj. As expected, the resolution is worse for more central events and for larger values of $R$, because of the larger UE contribution that must be subtracted. 
For $R \le 0.4 $ the difference found in  
both \pp and \PbPb simulations 
between the JES of 
generated and reconstructed \ptj is below  2\% at  mid-rapidity ($\abs{\eta} < 1$) and of order  4\% for ($1 < \abs{\eta} < 2$). 

A small nonclosure of up to 2\% was observed for all values of $R$ in the peripheral 70--100\% \PbPb bin, where the nonclosure is defined as the deviation of the corrected JES from unity.  
The UE in this bin is most comparable to that in \pp collisions, and it is used to evaluate the performance of the jet algorithm with heavy ion reconstruction and subtraction in the absence of UE.
This is necessary, as the difference in tracking and the subtraction of an UE in \PbPb, compared to \pp, results in modest performance changes even without a significant UE contribution.

The $\phi$-modulation of $\rho$ shown in 
Eq.~(\ref{eqn:PhiModulation}) improves the jet resolution without introducing any biases to the energy scale.  However, as can be seen in Fig.~\ref{fig:jesJer}, there is evidence of over-subtraction for the largest values of $R$ at low \ptj.  This is because of uncertainties in the  estimation of $\rho$. Errors in the estimation of $\rho$ tend to be handled much better for small $R$, as the  subtraction scales with the multiplicative area.
This over-subtraction causes the nonclosure to reach up to 4\% for $R = 1.0$ jets, as seen in Fig.~\ref{fig:jesJer}.
For smaller values of $R$ the nonclosure is below 2\%.

Another source of over-subtraction is caused by the flow-modulated subtraction. The minimum candidate count requirement for a good UE shape estimation does not account for the fact that jets could bias the fit.  The over-subtraction occurs when a jet biases the flow modulation fit.  While the fitting for $\phi$ modulation is turned off for events with a small number of tracks, events close to this threshold could still be affected by these biases, resulting in a source of nonclosure.

Finally, for the most central events, where $\rho(\eta, \phi)$ ranges from 200--300\GeV per unit area  and the fluctuations are the largest, there is a global underestimation of the true UE, particularly in the forward region ($\abs{\eta} > 1$). This occurs because towers within jets  are nominally excluded in the estimation of $\rho$ to avoid bias from the hard process.  In the most central events, upward fluctuations of the UE may cause some towers to be included in the jet and excluded from the UE. 
If too many towers are excluded, $\rho$ is underestimated. 
This underestimation of  $\rho$ results in the largest uncertainty in the final \RAA and spectra for the most central bins. It is mitigated by setting an upper limit on the number of excluded towers, with a cutoff that is tuned to achieve the best performance.

\subsection{Unfolding}

\begin{figure}[h!t]
\centering
       \includegraphics[width=0.75\textwidth]{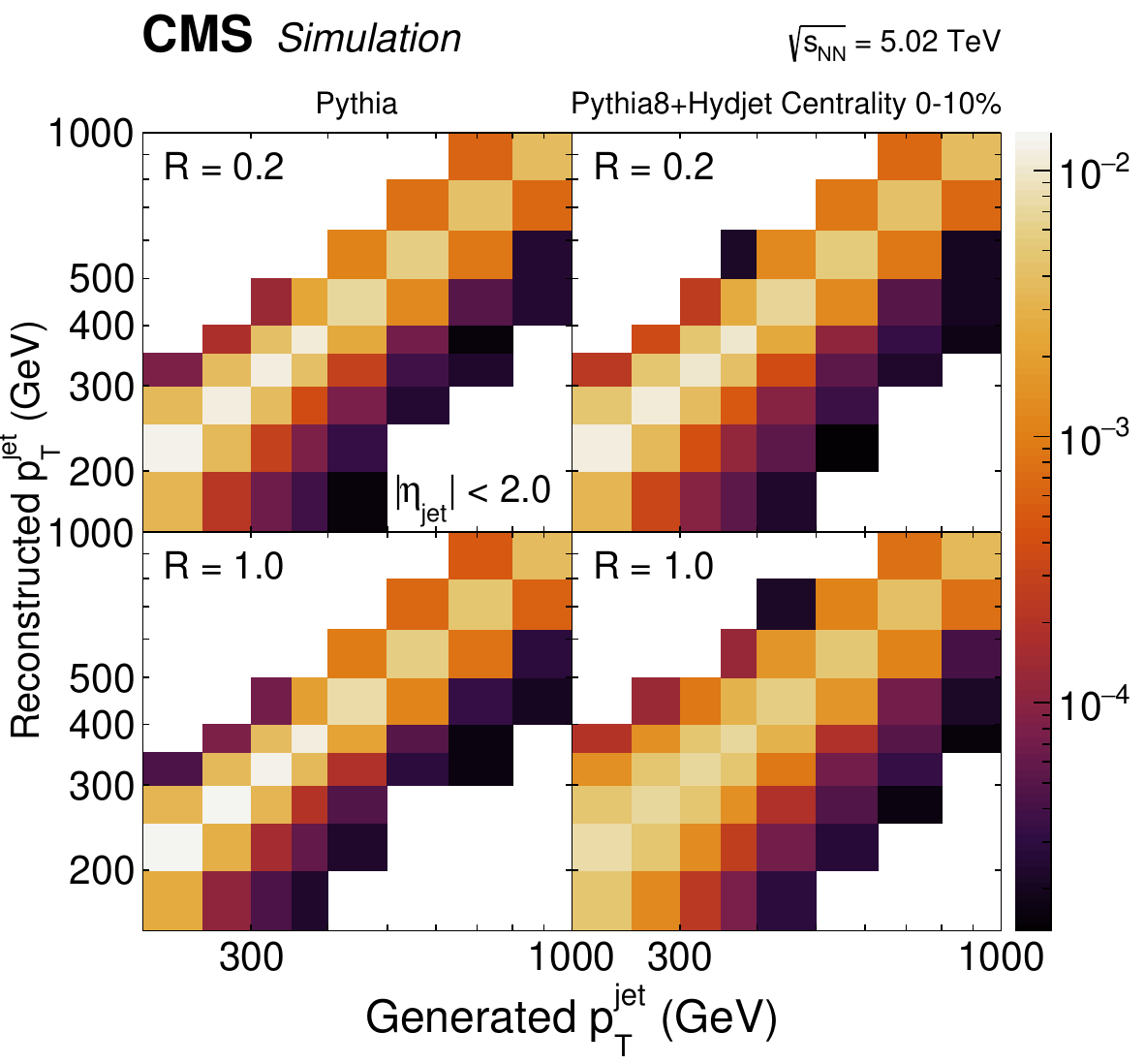}
       \caption{Response matrices in \PYTHIA (left) and \PYTHIAHYDJET 0--10\% \PbPb (right) events for jets reconstructed with $R=0.2$ (upper), $R=1.0$ (lower) and $\abs{\etaj}<2$.  The integral for each generated \ptj bin is normalized to unity.}
       \label{fig:R0p3Response}
\end{figure}

Raw spectra are unfolded according to response matrices constructed using \PYTHIAHYDJET MC for \PbPb and pure \PYTHIA for \pp results, in matched bins of \ptj, \etaj, and for \PbPb only,  event centrality.  The matrices are constructed with an {\ncoll} distribution that matches the expectations from data.
The unfolding is done with the d'Agostini iteration with early stopping~\cite{DAGOSTINI1995487} as implemented in the {\textsc{RooUnfold}} package~\cite{Adye:2011gm}.  Examples of response matrices are shown in Fig.~\ref{fig:R0p3Response} for \pp and 0--10\% \PbPb collisions with 
$R = 0.2$ and $1.0$.
Underflow bins are shown to account for bin migration. As expected, the matrices are more tightly  diagonal for \pp events than for \PbPb, and for $R = 0.2$ than for $R = 1.0$. 
The unfolded \pp and \PbPb spectra are then used to construct the \RAA distribution. 

\section{Systematic uncertainties}

The systematic uncertainties in the spectra are estimated by varying analysis parameters one at a time within a reasonable range, 
propagating 
the change through the full analysis chain, and then considering the deviation from the nominal results.
For \RAA, any correlation between the uncertainties in the \pp and \PbPb spectra is accounted for by simultaneously changing the same parameter in the \pp and \PbPb analyses, calculating a new \RAA and taking the difference from the nominal result.   
 This procedure produces a significant reduction in the uncertainty from data-simulation differences that impact JES and JER since the \pp and \PbPb were taken in run periods separated by just a few days. 
For ratios of \RAA between different jet radii, the luminosity and the \TAA uncertainties cancel. 

Finally, in the \RAA ratio between different radii, and the \pp ratios of spectra between radii, there are statistical cancellations as the same jet may contribute to multiple $R$ spectra. These are accounted for by comparing ratios of spectra in pseudo-experiments generated independently from the spectra and those generated with the correlation between different $R$ taken from the data.

\begin{figure}[h!t]
\centering
\includegraphics[width=0.96\textwidth]{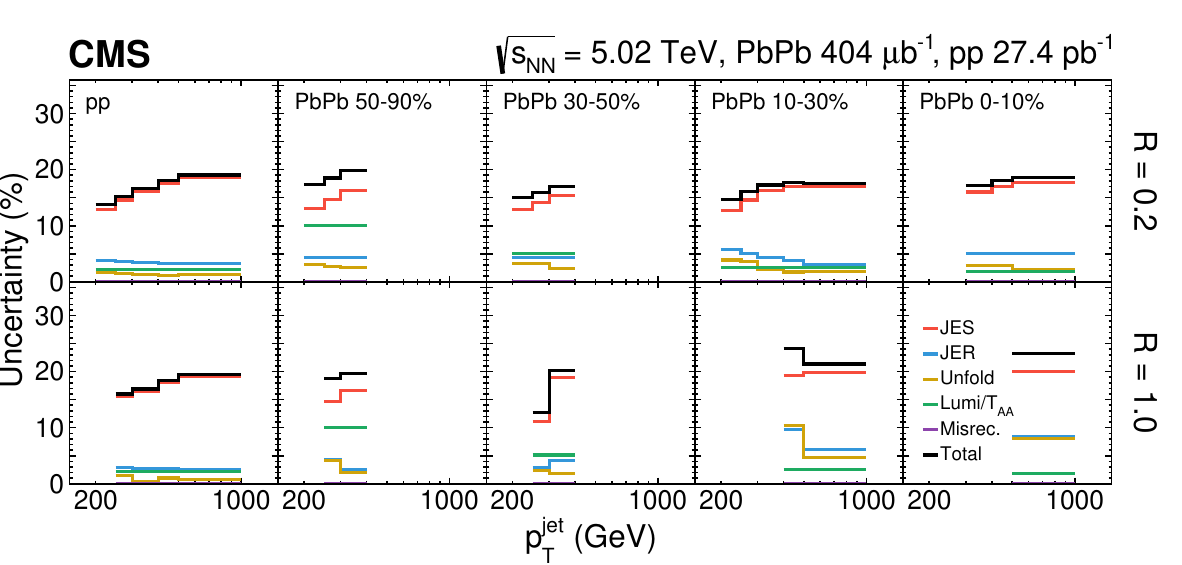}
\caption{Relative systematic uncertainties for the spectra of \ak jets within $\abs{\etaj} < 2.0$ for \pp collisions (left), and \PbPb with centrality classes 50--90\%, 30--50\%, 10--30\%, and 0--10\% (rightmost plot). The upper plots are for jets with $R=0.2$ and the lower plots for jets with $R=1.0$.}
\label{fig:systSpectra}
\end{figure}

Figure~\ref{fig:systSpectra} shows the principal systematic uncertainties as a function of \ptj for  $R=0.2$ and $1.0$, and for \pp and \PbPb collisions as a function of centrality. The dominant uncertainty arises from the JES. This tends to increase with \ptj and centrality but does not have a strong dependence upon $R$. The unfolding and JER uncertainties tend to decrease with \ptj and increase with centrality and $R$. The \TAA uncertainty decreases from peripheral to central events and is independent of \ptj and $R$. 

The origins of these uncertainties are listed below in order of importance for \RAA.  

\begin{enumerate}
   \item Jet energy scale.  The uncertainty ranges from 15 to 20\% and is dominated by the data-simulation difference.  It consists of several components, summed in quadrature:
      \begin{enumerate}
         \item Nonclosure in simulation.  This uncertainty is evaluated as a function of centrality and $\eta$ but independently of \ptj.  It is estimated by varying data by the observed nonclosure in simulation, see Fig.~\ref{fig:jesJer}, and then propagating this change through the analysis chain.  In \pp and peripheral (50--90\%) \PbPb collisions,
         a 1\% variation is made for all $\eta$.  For 10--50\% centrality, the variation is 1\% within $\abs{\etaj}<1$ and  2\% for $1<\abs{\etaj}<2$. For the most central (0--10\%) events, a 2\% variation is made for jets within $\abs{\etaj}<1$ and a 4\% variation for  $1<\abs{\etaj}<2$.
         \item Data-simulation differences.  A flat 2\% variation is performed in all bins following the procedure used for the nonclosure uncertainties above. This uncertainty is dominant in the \pp spectra, and  comparable to  the nonclosure uncertainty in semicentral and semiperipheral \PbPb bins.
         \item Differences from the UE description between data and simulation. These differences are extracted by comparing random cone mean/widths between data and simulation, and the full difference is taken as a systematic uncertainty. As this is the centrality-dependent component of the JES, it does not cancel in the ratios between \pp and \PbPb data, and only cancels partially in $R$-dependent ratios of \RAA.
      \end{enumerate}
   \item Jet energy resolution.
         \begin{enumerate}
         \item The JER uncertainty is extracted from simulation. This is subdominant compared to the data-simulation differences for spectra, but does not cancel in \RAA.
       \item Jet energy resolution from data-simulation differences.  The resolution in data is found to be 10 to 15\% worse than that in simulation. To propagate this uncertainty, the simulation is first smeared by 10\%, such that central values are closer to those in data. The systematic uncertainty is estimated by applying an additional smearing on top of these new central values such that the resolution is increased by 10\% in all bins.  
      The effect is subdominant in part because the 
      \ptj binning was chosen to minimize bin migration. Furthermore, there is partial cancellation in \RAA, coming from the constant and stochastic terms of the resolution, which are partially shared between the \pp and \PbPb data.
      \end{enumerate}
   \item Unfolding.  This source of uncertainty is typically of order 5\% with a maximum of 10\%.  
   There are several components within this category:
      \begin{enumerate}
         \item The choice of the prior.  A variation of the nominal prior for the underlying \ptj spectrum is done and propagated through the full analysis chain, including the response matrix.  
         \item Unfolding algorithm.  The result is cross-checked with singular value decomposition unfolding~\cite{SVDUnfolding}.
      \end{enumerate}
   \item Integrated luminosity and \TAA.  The uncertainty in the integrated luminosity for \pp collisions is 2.3\%~\cite{CMS-PAS-LUM-16-001}. For the \TAA, the relative uncertainties vary between 3\% for the 0--10\% bin, to 11\% in the most peripheral 50--90\% bin~\cite{Loizides:2017ack}. The absolute uncertainties for each of the four values are listed in Table~\ref{tab:TAA}.
     \item Misreconstructed jets which arise from fluctuations of the UE.  The contamination from these jets is evaluated from simulation, and it is found to be negligible in the considered kinematic range.
\end{enumerate}

\section{Results}

\begin{figure}[h!t]
\centering
        \includegraphics[width=0.95\textwidth]{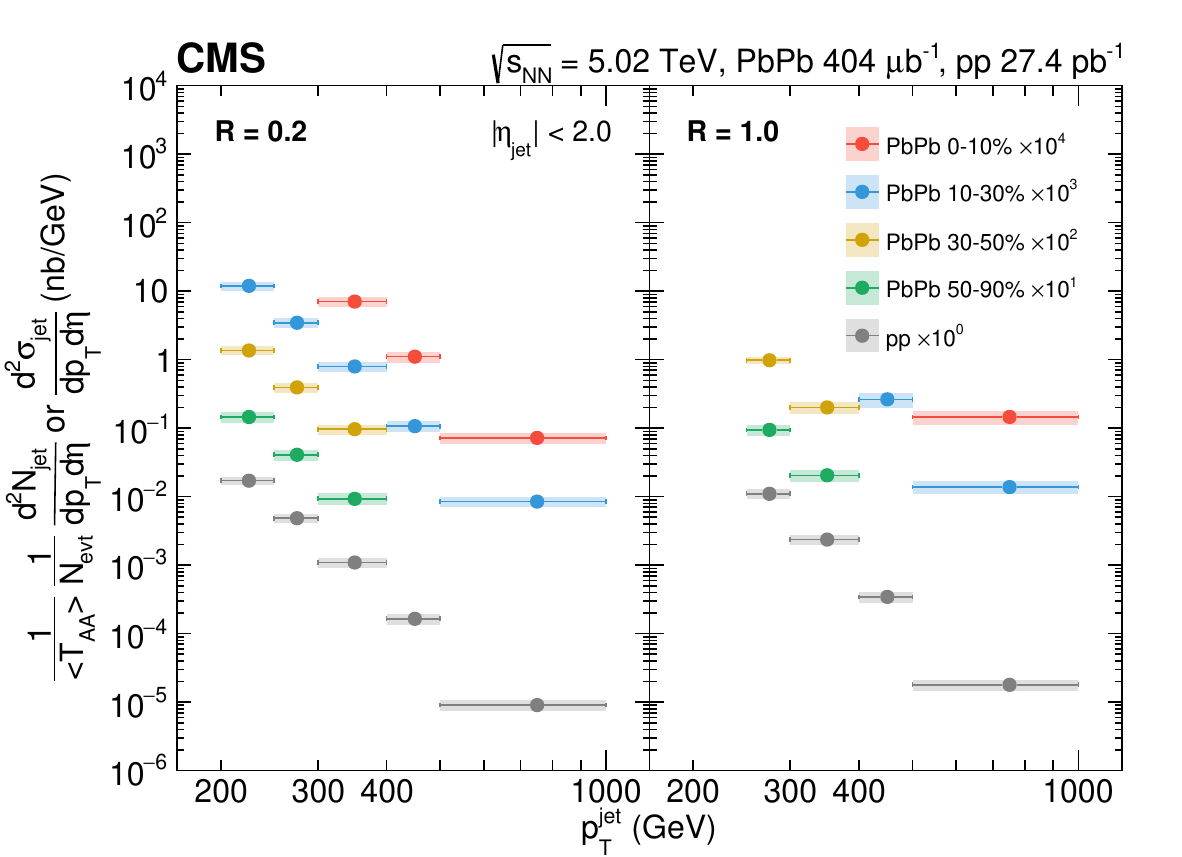}
       \caption{Spectra of jets with $\abs{\etaj} < 2.0$ for $R=0.2$ (left) and $R=1.0$ (right), for \pp collisions and different centrality classes of \PbPb collisions.  
       The spectra are multiplied by successive factors of 10 for clarity. 
       The statistical uncertainties are smaller than the marker sizes, while the systematic uncertainties are  shown as shaded boxes.  The markers are placed at the bin centers.}
       \label{fig:resultsSpectra}
\end{figure}

The unfolded jet spectra as functions of \ptj for $R=0.2$ and $1.0$ for both \pp and \PbPb collisions of various centralities are shown in Fig.~\ref{fig:resultsSpectra}.
The lower bound of \ptj is chosen based on the observed noise level for each centrality class, and the upper bound is driven by the amount of statistics.

\begin{figure}[h!t]
\centering
       \includegraphics[width=0.6\textwidth]{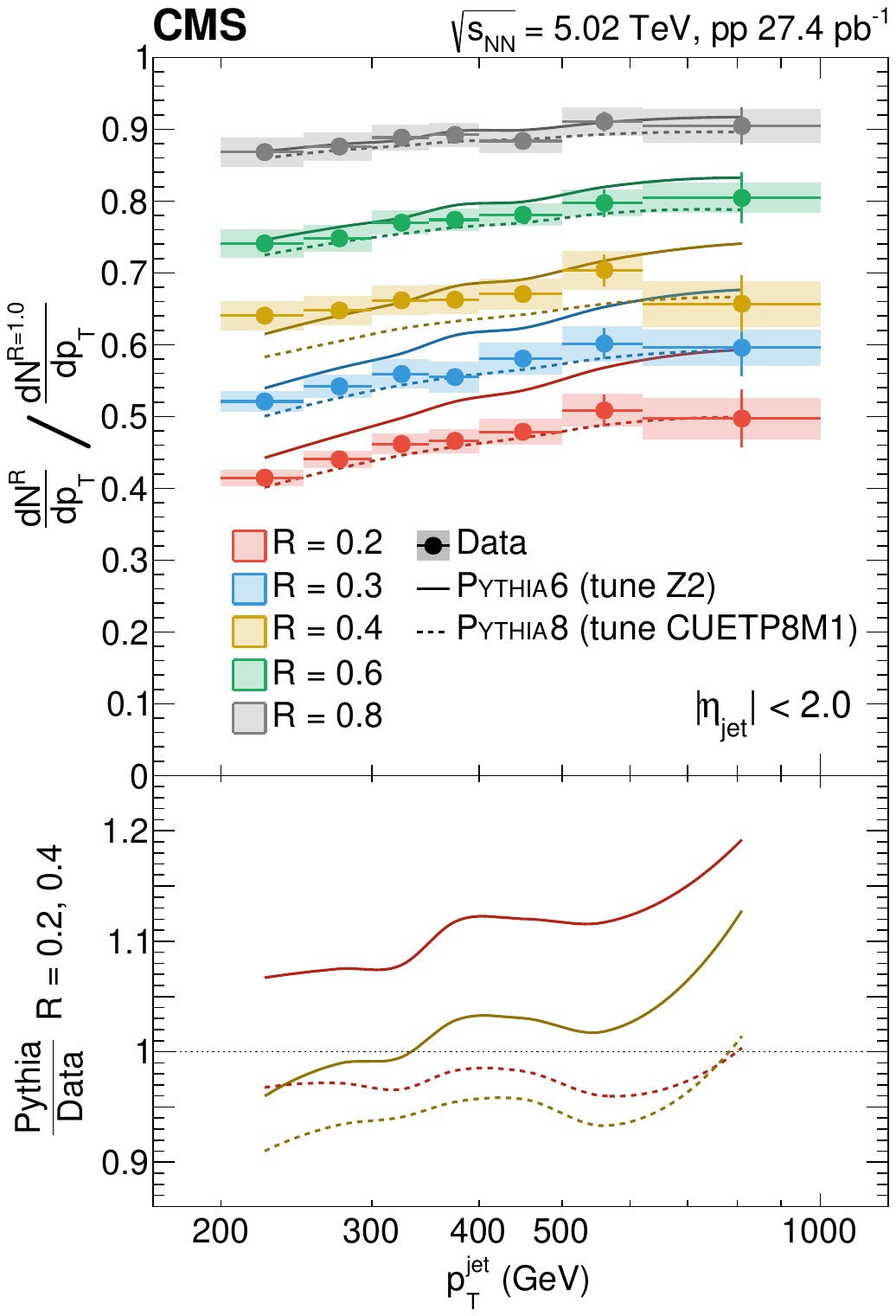}
       \caption{The spectra ratio for jets from \pp collisions with $\abs{\etaj} < 2.0$ for $R=0.2$--$0.8$ with respect to $R=1.0$.  The statistical uncertainty of data is shown as vertical lines, whereas the systematic uncertainties are shown as the shaded boxes.  Markers for the data are placed at the bin centers.  Comparisons with \PYTHIA{6} (solid line) and \PYTHIA{8} (dotted line) are plotted, along with ratios in the lower plot for $R=0.2$ and $R=0.4$.}
       \label{fig:resultsSpectRat}
\end{figure}

The upper plot of Fig.~\ref{fig:resultsSpectRat} shows the ratio of spectra of jets with different radii in \pp collisions, normalized to the spectrum for $R = 1.0$. The number of jets with a given \ptj increases with the size of the jet cone.
The increase of jet yield with $R$ becomes weaker at higher values of \ptj  suggesting that jets become narrower as \ptj increases.
Figure~\ref{fig:resultsSpectRat} also shows predictions using the \PYTHIA{6}  and \PYTHIA{8} MC generators.  Both generators capture the trends of the data but  \PYTHIA{8} is closer to the scale of the data.  The lower plot of Fig.~\ref{fig:resultsSpectRat} shows the ratios of the jet spectra from \PYTHIA to the data spectrum for $R=0.2$ and $0.4$. For \PYTHIA{6} the ratio rises with \ptj for both values of $R$. The  \PYTHIA{8} ratios show little dependence on \ptj and are generally closer to unity than those of \PYTHIA{6}. 

\begin{figure}[htb]
\centering
      \includegraphics[width=0.96\textwidth]{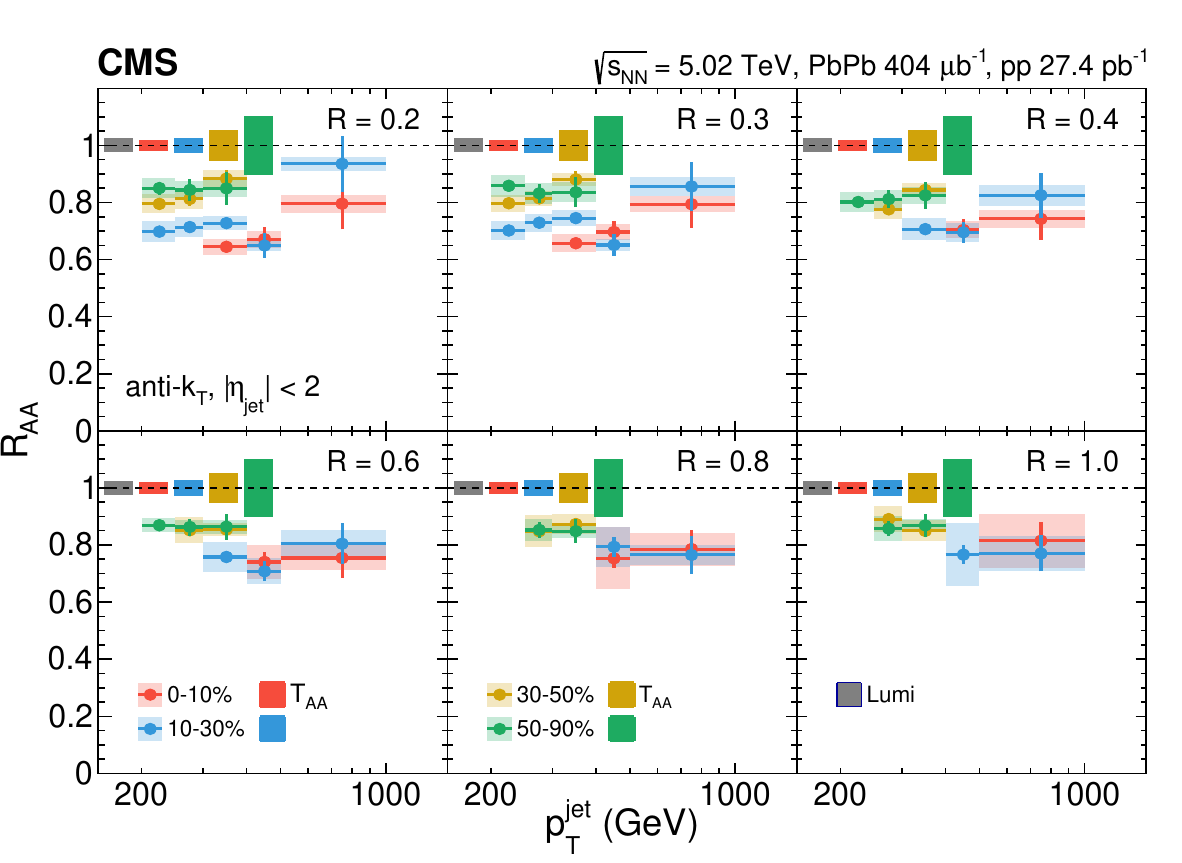}
      \caption{The \RAA for jets with $\abs{\etaj} < 2.0$ as functions of \ptj for various $R$ and centrality classes.  The statistical uncertainties are represented by  vertical lines, and the systematic uncertainties by shaded boxes.  The markers are placed at the bin centers.  Global uncertainties (integrated luminosity for \pp and \TAA for PbPb data) are shown as colored boxes on the dashed line at $\RAA = 1$ and are not included in the shaded boxes around the points.}
       \label{fig:resultsRAA}
\end{figure}

The \RAA factors compare PbPb data to the scaled \pp reference.
Figure~\ref{fig:resultsRAA} shows \RAA, the ratio of \PbPb data to a scaled \pp reference, as functions of \ptj, jet radius, and centrality. Systematic uncertainties related to the JES and JER cancel partially. The remaining systematic uncertainties are dominated by the uncertainties in the integrated luminosity, \TAA, and the JES uncertainty component from the UE.

For all values of $R$, \RAA for the most peripheral collisions (50--90\%) is independent of \ptj and consistent with unity after considering the \TAA uncertainty. In the most central bin, a strong suppression of the \PbPb data ($\approx$0.6--0.7) is observed, which is well outside the systematic uncertainties.
However, there are hints of an increasing \RAA with \ptj for the smaller values of $R$ in the central bins, with values up to 0.8 for jets with $\ptj > 500\GeV$.

\begin{figure}[htb]
        \centering
      \includegraphics[width=1.0\textwidth]{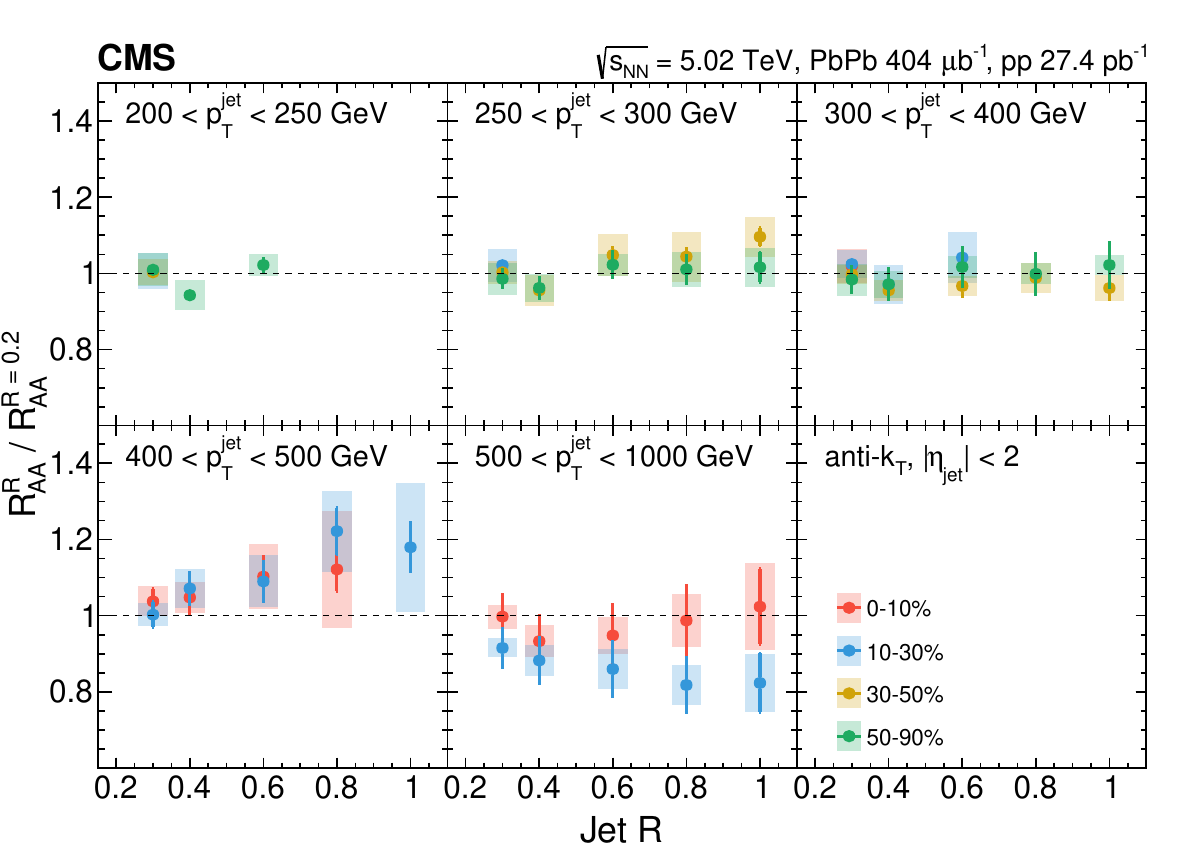}
      \caption{The \RAA ratio for jets with $\abs{\etaj} < 2.0$ as a function of $R$ for $R=0.3$--$1.0$ with respect to $R=0.2$, in various event centrality classes and \ptj ranges.  The statistical uncertainties of data are shown as the vertical lines, whereas the systematic uncertainties are shown as the shaded boxes.}
       \label{fig:resultsRRAA}
\end{figure}

To highlight the jet radius dependence of the jet \RAA, the ratios of \RAA 
for a given $R$ with respect to $R=0.2$ are presented in Fig.~\ref{fig:resultsRRAA}.
This observable is particularly sensitive to the recovery of the quenched energy and the presence of the medium response~\cite{pablos2019jet}. 
For $400 < \ptj < 500 \GeV$, the \RAA ratios are above unity and increase with \ptj in both the 0--10\% and 10--30\% centrality intervals. 
On the other hand, for $\ptj > 500 \GeV$, the \RAARatio is close to unity or slightly below it  for the 
0--10\% and 10--30\% centrality intervals, respectively.

\begin{figure}[h]
\centering
      \includegraphics[width=1.0\textwidth]{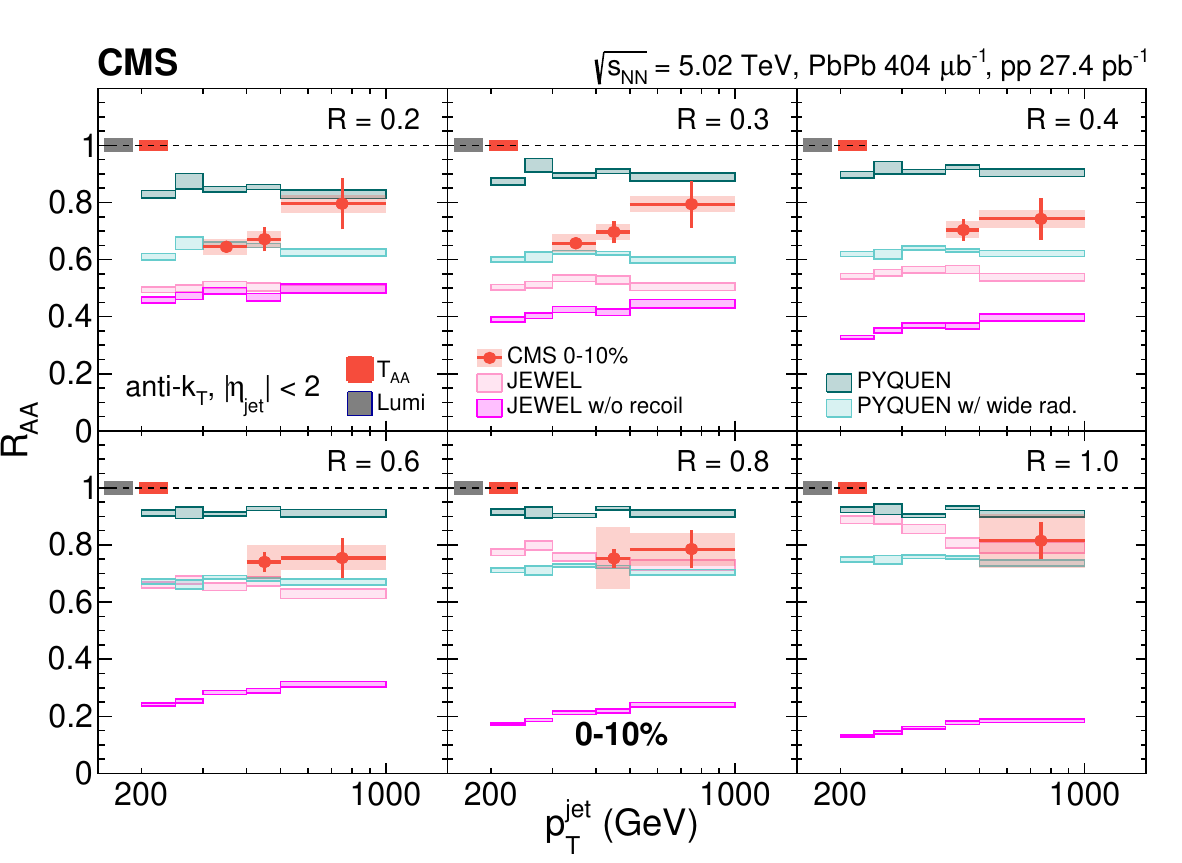}
      \caption{The \RAA for jets with $\abs{\etaj} < 2.0$, as a function of \ptj, for various $R$ and 0--10\% centrality class.  The statistical uncertainties are represented by the vertical lines, while the systematic uncertainties are shown as the shaded boxes.  The markers are placed at the bin centers.  Global uncertainties (integrated luminosity for \pp and \TAA for PbPb collisions) are shown as the colored boxes on the dashed line at $\RAA = 1$ and are not included in the shaded bands around the points.  The predictions from \jewel (fuchsia and pink) and \pyquen (teal and turquoise) generators, shown as colored boxes, are compared to the data.}
       \label{fig:resultsRAAGen}
\end{figure}

Figure~\ref{fig:resultsRAAGen} shows  \RAA 
 for 0--10\% central \PbPb collisions, as a function of \ptj for several $R$ values.  As \ptj increases, \RAA increases. 
Also shown in  Fig.~\ref{fig:resultsRAAGen} are 
predictions from the  \jewel~\cite{Zapp2014JEWEL2D} (v2.2.0) and \pyquen~\cite{Lokhtin:2005px} (v1.5.4) generators.
The \jewel predictions are made 
with (pink) and without (fuchsia) contributions from recoil particles (\ie scattered medium particles). The predictions without recoil particles are in disagreement with the data, showing the importance of the medium response.  The importance of recoil particles within \jewel increases greatly as $R$ increases. For $R=1.0$
the predictions without recoil are a factor of four below the default mode with recoil.
The \jewel predictions with recoil are significantly below the data for $R=0.2$ but come increasingly close to the data as $R$ increases.  

Predictions from \pyquen  are shown 
 with (the default, shown in teal) and without (turquoise) medium-induced wide-angle radiation.
The default \pyquen generator  overpredicts \RAA particularly  for smaller values of  $R$ and \ptj.  The inclusion of wide-angle radiation lowers the predictions for \RAA particularly for smaller $R$ sizes and brings the \pyquen predictions closer to the data, showing the importance of the medium effects.

\begin{figure}[htb]
        \centering
      \includegraphics[width=0.76\textwidth]{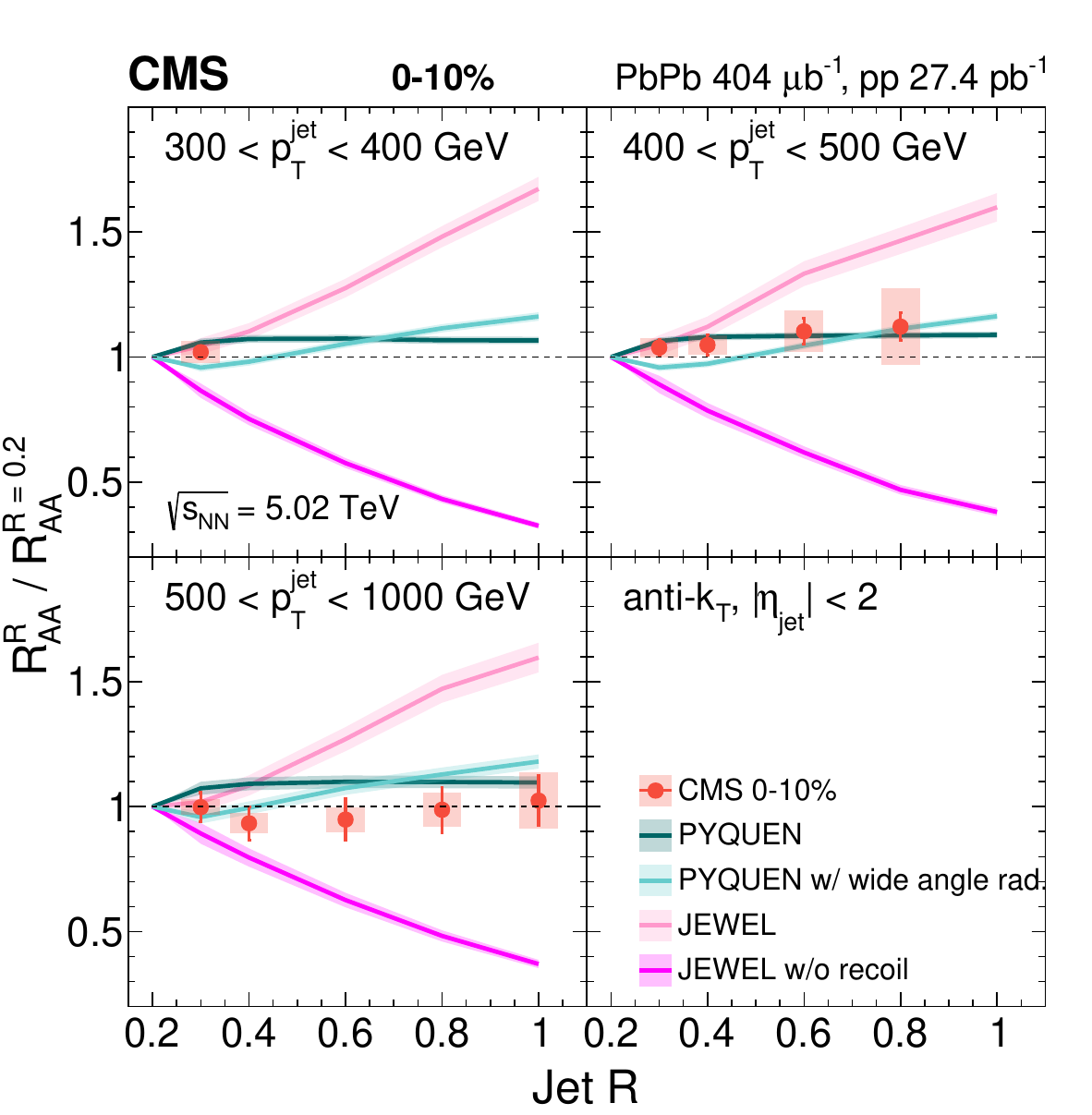}
      \caption{The \RAA ratio for jets with $\abs{\etaj} < 2.0$ as a function of $R$ for $R=0.3$--$1.0$ with respect to $R=0.2$, in various \ptj ranges for the 0--10\% centrality class.  The statistical uncertainties of data are shown as the vertical lines, whereas the systematic uncertainties are shown as the shaded boxes.  The width of the boxes carries no meaning.  The predictions from \jewel (fuchsia and pink) and \pyquen (teal and turquoise) generators, shown with the colored bands, are compared to the data.}
       \label{fig:resultsRRAAGen}
\end{figure}

Figure~\ref{fig:resultsRRAAGen} shows \RAARatio  as a function of $R$ for several values of  \ptj. Monte Carlo predictions from the  \jewel and \pyquen  generators are also shown. 
For  the data,  \RAARatio 
has little dependence upon $R$ and is consistent with unity for all values of \ptj for both the data and the \pyquen predictions.  The \jewel  model is unable to capture the $R$ dependence of \RAARatio. For the 
predictions with recoil, \RAARatio  increases as a function of $R$ but if  recoil is ignored \RAARatio decreases with $R$.

\begin{figure}[h!t]
\centering
      \includegraphics[width=0.96\textwidth]{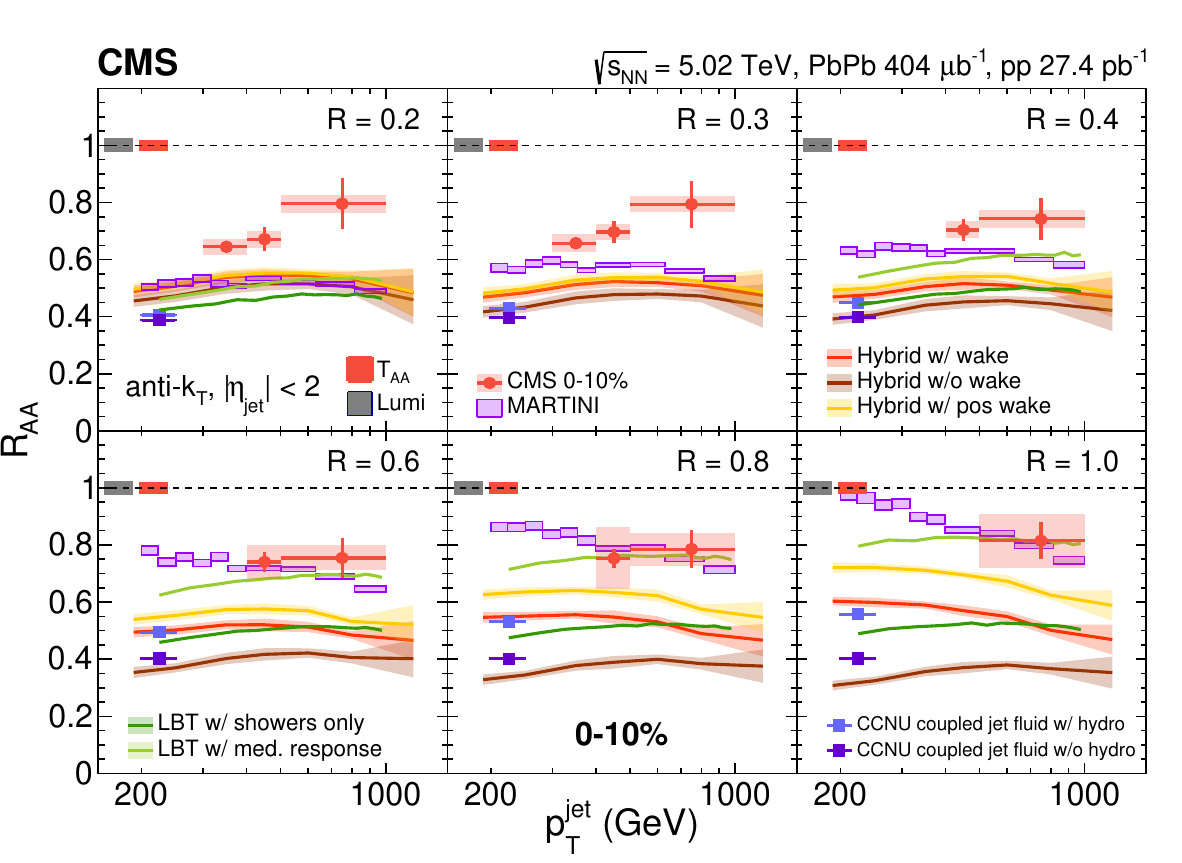}
      \caption{The \RAA for jets with $\abs{\etaj} < 2.0$, as a function of \ptj, for various $R$ values and the 0--10\% centrality class.  The statistical uncertainties are represented by  vertical lines, while the systematic uncertainties are shown as  shaded boxes.  The markers are placed at the bin centers.  Global uncertainties (integrated luminosity for \pp and \TAA for PbPb collisions) are shown as the colored boxes on the dashed line at \RAA = 1 and  are not included in the shaded bands around the points.  The predictions from \Hybrid (dark orange, brown and yellow), \Martini (purple), \LBT (lime and dark green), and \CCNU (blue and violet) models, shown as the colored boxes and bands, are compared to the data.}
       \label{fig:resultsRAAMC}
\end{figure}

Figure~\ref{fig:resultsRAAMC} shows a comparison of several models to \RAA  as functions of \ptj and $R$. 
The  \Hybrid model~\cite{pablos2019jet} combines a perturbative description of the weakly coupled physics of jet production and evolution, 
with a gauge/gravity duality description of the strongly coupled dynamics of the medium, and the soft-gluon exchanges between the jet and medium. 
As the jet passes through and deposits energy into the hydrodynamic medium, a wake is left behind the jet. 
The \Hybrid model (dark orange) tends to under-predict \RAA at high \ptj.  
Calculations without a wake (brown) and with only the positive contribution of the wake (yellow) are also shown.
These two are not physical and are included here only for better understanding of the effect of the wake contribution.
The effect of the wake is more important at large $R$ and lower \ptj.

In the Linear Boltzmann Transport (\LBT) model~\cite{He:2018xjv}, 
the effects of recoil thermal partons and their  propagation in the dense medium are described by a 3+1D viscous relativistic hydrodynamic model.  Predictions from \LBT are shown in 
 Fig.~\ref{fig:resultsRAAMC} with and without the medium response. It is clear that the medium response becomes more and more dominant as the size of the jet cone increases. 
  A similar effect is seen for the  jet-coupled fluid model~\cite{Chang:2016gjp,Tachibana:2017syd,Chang:2019sae} \CCNU. Although predictions are only available for a limited \ptj range, it is clear from comparing the blue and violet points in  Fig.~\ref{fig:resultsRAAMC} that the hydrodynamic component of \CCNU  becomes increasingly important with increasing $R$.

The predictions from \Martini~\cite{Schenke:2009gb} (Modular Algorithm for Relativistic Treatment of Heavy IoN Interactions) are shown as purple boxes in Fig.~\ref{fig:resultsRAAMC}. 
The model follows a hybrid approach where it embeds the high energy parton into an evolving hydrodynamic medium, and the shower evolution of the jet is modified following the McGill-AMY formalism~\cite{McGillAMY1,McGillAMY2,McGillAMY3,McGillAMY4,McGillAMY5}.
The \Martini generator predicts a larger increase of jet \RAA ratio as a function of $R$ than what is observed in data. 

From Fig.~\ref{fig:resultsRAAMC},  
it is striking that, for the small jet radius $R=0.2$, \RAA rises with \ptj but  the 
\Hybrid, \LBT, \CCNU and \Martini models  are all flat in \ptj.  For all these 
models, hydrodynamic or medium effects become increasingly important as $R$ increases and are indeed dominant for $R=1.0$. 

\begin{figure}[h!t]
        \centering
      \includegraphics[width=0.76\textwidth]{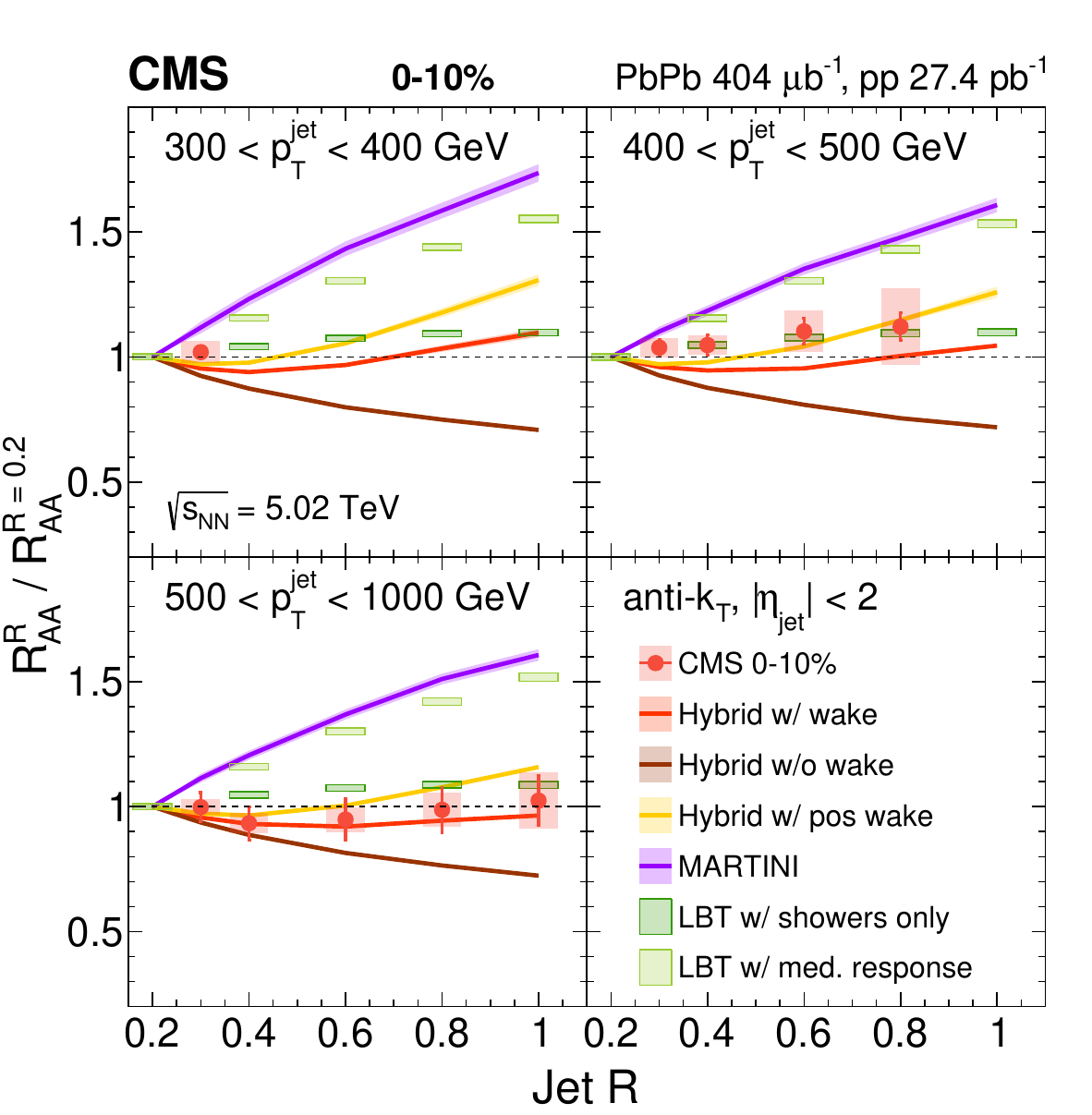}
      \caption{The double ratio \RAARatio for jets with $\abs{\etaj} < 2.0$, as a function of $R$, for $R=0.3$--$1.0$ with respect to $R=0.2$, in various \ptj ranges for the 0--10\% centrality class.   The statistical uncertainties of data are shown as the vertical lines, whereas the systematic uncertainties are shown as the shaded boxes.  The width of the boxes carries no meaning.  The predictions from the \Hybrid (dark orange, brown and yellow), \Martini (purple), and \LBT (lime and dark green) models are compared to the data as colored bands.}
       \label{fig:resultsRRAAMC}
\end{figure}

Figure~\ref{fig:resultsRRAAMC} 
shows \RAARatio  as a function of $R$, for several values of  \ptj. Monte Carlo predictions from the  \Hybrid, \Martini,  and \LBT generators are also shown.  The \Hybrid model (orange) is able to describe the data.
However if the wake contribution is ignored (brown) the model gives a different trend.
 The \Martini model (purple) predicts that \RAARatio should increase with $R$ in contrast to the data. The default \LBT model (lime)  is consistent with the data but \LBT with showers only and no medium response (dark green) overpredicts \RAARatio.
 Some of the models which correctly predict the trend of \RAARatio are off in the \RAA, as can be seen in Fig.~\ref{fig:resultsRAAMC}.

\begin{figure}[h!t]
\centering
      \includegraphics[width=1.00\textwidth]{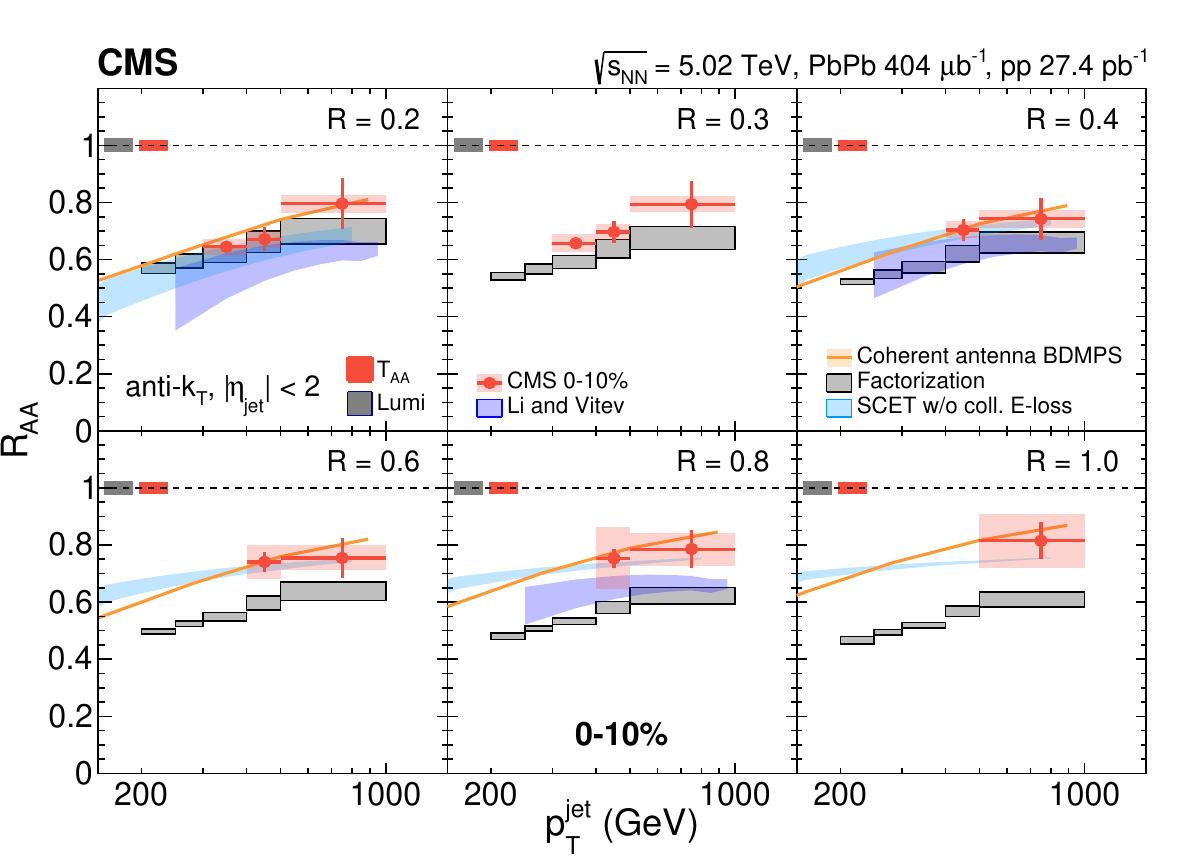}
      \caption{The \RAA for jets with $\abs{\etaj} < 2.0$, as a function of \ptj, for various $R$ values and 0--10\% centrality class.  The statistical uncertainties are represented by the vertical lines, while the systematic uncertainties are shown as the shaded boxes.  The markers are placed at the bin centers.  Global uncertainties (integrated luminosity for \pp and \TAA for PbPb collisions) are shown as the colored boxes on the dashed line at $\RAA = 1$ and  are not included in the shaded bands around the points.  The calculations from \SCET  (sky blue and navy blue), coherent antenna \BDMPS (orange) and jet factorization (gray) formalisms are compared to the data, shown as the colored boxes and bands.}
       \label{fig:resultsRAACalc}
\end{figure}

The same data in Figs.~\ref{fig:resultsRAAMC} and~\ref{fig:resultsRRAAMC} are also compared to additional models.
Figure~\ref{fig:resultsRAACalc} shows \RAA vs. \ptj for several values of $R$ and for the top 0--10\% centrality class as well as several predictions from generators and analytic calculations.
The gray boxes in Fig.~\ref{fig:resultsRAACalc} are predictions from a jet factorization model based on a phenomenological approach to establish QCD factorization of jet cross sections in heavy ion collisions~\cite{Qiu_2019}. 
Medium-modified jet functions are extracted from jet nuclear modification factors at smaller jet distance parameter values ($R=0.2$ and 0.4) and predictions are made for larger distance parameter values.
At $R<0.4$, the data are described reasonably well by the factorization model.  
However, the model tends to underpredict \RAA at larger $R$ values.
The data are also compared to the coherent antenna \BDMPS calculations~\cite{Baier:1994bd} (orange), which is an analytical approach that resums multiple emissions to leading-logarithmic accuracy including both radiative energy loss and color coherence effects~\cite{CasalderreySolana:2012ef,Mehtar-Tani:2016aco,Mehtar-Tani:2017web}. 
The predictions are in general agreement with the \RAA  data.

Finally, calculations based on a soft collinear effective theory with Glauber gluon interactions \SCET~\cite{Chien:2015hda}, are also compared to the  data. 
The \SCET calculations with collisional energy loss~\cite{Li2019,Sievert_2019} (navy blue) 
are slightly below the \RAA measurements while those without collisional energy loss (sky blue) are consistent with the data.

\begin{figure}[h!t]
        \centering
      \includegraphics[width=0.76\textwidth]{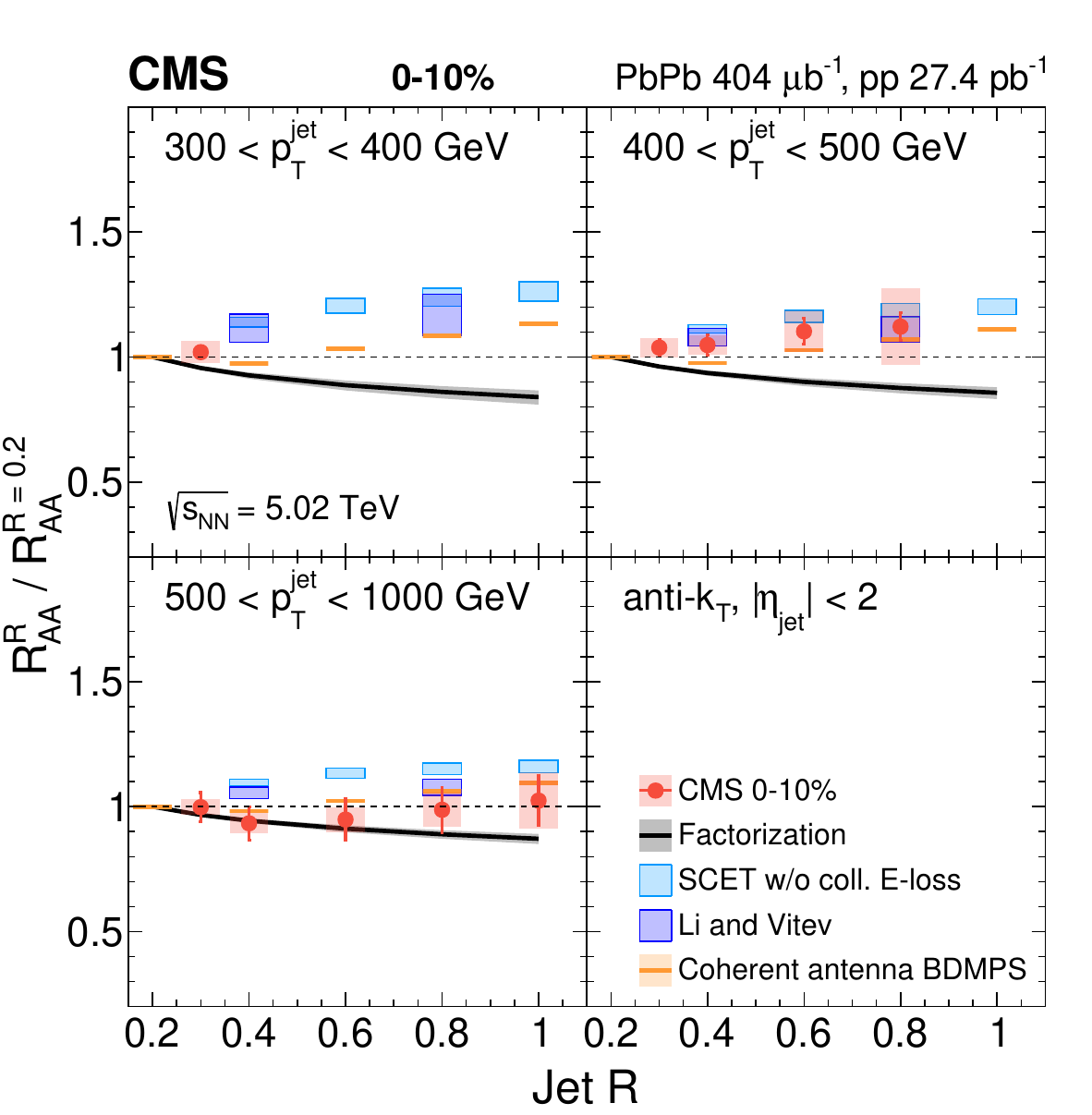}
      \caption{The double ratio \RAARatio for jets with $\abs{\etaj} < 2.0$ as a function of $R$ for $R=0.3$--$1.0$ with respect to $R=0.2$, in various \ptj ranges for the 0--10\% centrality class.   The statistical uncertainties of data are shown as the vertical lines, whereas the systematic uncertainties are shown as the shaded boxes.  The width of the boxes carries no meaning.  The calculations based from \SCET  (sky blue and navy blue), coherent antenna \BDMPS (orange) and, jet factorization (gray) formalisms, shown with the colored bands and boxes, are compared to the data.}
       \label{fig:resultsRRAACalc}
\end{figure}

Figure~\ref{fig:resultsRRAACalc} shows \RAARatio vs. $R$ for several  values of \ptj together with predictions from the 
\SCET, \BDMPS and jet factorization models.  The \BDMPS (orange) and \SCET predictions (sky blue and navy blue) are consistent with the data but the factorization calculations (gray) decrease with $R$ in contrast to the data.

\section{Summary}

Measurements of jet nuclear modification factors based on proton-proton and lead-lead collisions at $\sqrtsNN = 5.02\TeV$ are presented. For the first time, jet spectra measurements are extended to large area jets, with a \ak distance parameter $R$ up to 1.0. 
For the most central \PbPb collisions, a strong suppression is observed for jets with high transverse momentum reconstructed with all distance parameters. 
Predictions from quenched jet event generators, theoretical models, and analytical calculations are compared to these results.
The new data place further constraints on the underlying jet quenching mechanisms.
While state of the art models have made important progress, significant tension remains in view of the large area jet data presented here.

\clearpage

\begin{acknowledgments}
   We congratulate our colleagues in the CERN accelerator departments for the excellent performance of the LHC and thank the technical and administrative staffs at CERN and at other CMS institutes for their contributions to the success of the CMS effort. In addition, we gratefully acknowledge the computing  centers and personnel of the Worldwide LHC Computing Grid and other  centers for delivering so effectively the computing infrastructure essential to our analyses. Finally, we acknowledge the enduring support for the construction and operation of the LHC, the CMS detector, and the supporting computing infrastructure provided by the following funding agencies: BMBWF and FWF (Austria); FNRS and FWO (Belgium); CNPq, CAPES, FAPERJ, FAPERGS, and FAPESP (Brazil); MES (Bulgaria); CERN; CAS, MoST, and NSFC (China); COLCIENCIAS (Colombia); MSES and CSF (Croatia); RIF (Cyprus); SENESCYT (Ecuador); MoER, ERC PUT and ERDF (Estonia); Academy of Finland, MEC, and HIP (Finland); CEA and CNRS/IN2P3 (France); BMBF, DFG, and HGF (Germany); GSRT (Greece); NKFIA (Hungary); DAE and DST (India); IPM (Iran); SFI (Ireland); INFN (Italy); MSIP and NRF (Republic of Korea); MES (Latvia); LAS (Lithuania); MOE and UM (Malaysia); BUAP, CINVESTAV, CONACYT, LNS, SEP, and UASLP-FAI (Mexico); MOS (Montenegro); MBIE (New Zealand); PAEC (Pakistan); MSHE and NSC (Poland); FCT (Portugal); JINR (Dubna); MON, RosAtom, RAS, RFBR, and NRC KI (Russia); MESTD (Serbia); SEIDI, CPAN, PCTI, and FEDER (Spain); MOSTR (Sri Lanka); Swiss Funding Agencies (Switzerland); MST (Taipei); ThEPCenter, IPST, STAR, and NSTDA (Thailand); TUBITAK and TAEK (Turkey); NASU (Ukraine); STFC (United Kingdom); DOE and NSF (USA).
    
   \hyphenation{Rachada-pisek} Individuals have received support from the Marie-Curie program and the European Research Council and Horizon 2020 Grant, contract Nos.\ 675440, 724704, 752730, and 765710 (European Union); the Leventis Foundation; the Alfred P.\ Sloan Foundation; the Alexander von Humboldt Foundation; the Belgian Federal Science Policy Office; the Fonds pour la Formation \`a la Recherche dans l'Industrie et dans l'Agriculture (FRIA-Belgium); the Agentschap voor Innovatie door Wetenschap en Technologie (IWT-Belgium); the F.R.S.-FNRS and FWO (Belgium) under the ``Excellence of Science -- EOS" -- be.h project n.\ 30820817; the Beijing Municipal Science \& Technology Commission, No. Z191100007219010; the Ministry of Education, Youth and Sports (MEYS) of the Czech Republic; the Deutsche Forschungsgemeinschaft (DFG), under Germany's Excellence Strategy -- EXC 2121 ``Quantum Universe" -- 390833306, and under project number 400140256 - GRK2497; the Lend\"ulet (``Momentum") Program and the J\'anos Bolyai Research Scholarship of the Hungarian Academy of Sciences, the New National Excellence Program \'UNKP, the NKFIA research grants 123842, 123959, 124845, 124850, 125105, 128713, 128786, and 129058 (Hungary); the Council of Science and Industrial Research, India; the Ministry of Science and Higher Education and the National Science Center, contracts Opus 2014/15/B/ST2/03998 and 2015/19/B/ST2/02861 (Poland); the National Priorities Research Program by Qatar National Research Fund; the Ministry of Science and Higher Education, project no. 0723-2020-0041 (Russia); the Programa Estatal de Fomento de la Investigaci{\'o}n Cient{\'i}fica y T{\'e}cnica de Excelencia Mar\'{\i}a de Maeztu, grant MDM-2015-0509 and the Programa Severo Ochoa del Principado de Asturias; the Thalis and Aristeia programs cofinanced by EU-ESF and the Greek NSRF; the Rachadapisek Sompot Fund for Postdoctoral Fellowship, Chulalongkorn University and the Chulalongkorn Academic into Its 2nd Century Project Advancement Project (Thailand); the Kavli Foundation; the Nvidia Corporation; the SuperMicro Corporation; the Welch Foundation, contract C-1845; and the Weston Havens Foundation (USA). 
\end{acknowledgments}

\bibliography{auto_generated}

\cleardoublepage \appendix\section{The CMS Collaboration \label{app:collab}}\begin{sloppypar}\hyphenpenalty=5000\widowpenalty=500\clubpenalty=5000\vskip\cmsinstskip
\textbf{Yerevan Physics Institute, Yerevan, Armenia}\\*[0pt]
A.M.~Sirunyan$^{\textrm{\dag}}$, A.~Tumasyan
\vskip\cmsinstskip
\textbf{Institut f\"{u}r Hochenergiephysik, Wien, Austria}\\*[0pt]
W.~Adam, F.~Ambrogi, T.~Bergauer, M.~Dragicevic, J.~Er\"{o}, A.~Escalante~Del~Valle, M.~Flechl, R.~Fr\"{u}hwirth\cmsAuthorMark{1}, M.~Jeitler\cmsAuthorMark{1}, N.~Krammer, I.~Kr\"{a}tschmer, D.~Liko, T.~Madlener, I.~Mikulec, N.~Rad, J.~Schieck\cmsAuthorMark{1}, R.~Sch\"{o}fbeck, M.~Spanring, W.~Waltenberger, C.-E.~Wulz\cmsAuthorMark{1}, M.~Zarucki
% ...
\vskip\cmsinstskip
\dag: Deceased\\
1:  Also at Vienna University of Technology, Vienna, Austria\\
% ...
\end{sloppypar}
\end{document}